\documentclass[prd,twocolumn,superscriptaddress,amsfonts,amssymb,amsmath,showpacs]{revtex4-2}
\usepackage{bm}
\usepackage{latexsym}
\usepackage[latin1]{inputenc}
\usepackage{graphicx}
\usepackage{amsmath}
\usepackage{amssymb}
\usepackage{amsthm}
\usepackage{relsize}
\usepackage{palatino}
\usepackage{mathpazo}
\usepackage{textcomp}
\usepackage{float}
\usepackage{booktabs}
\usepackage{dcolumn}
\usepackage{booktabs}
\usepackage{multirow}
\usepackage{tikz}
\usepackage{hyperref}
\hypersetup{
    colorlinks=true,
    linkcolor=purple,
    filecolor=magenta,      
    citecolor=blue
}
\usepackage{amsmath}
\usepackage{xcolor}
\usepackage{orcidlink}
\usepackage[caption=false]{subfig}
\usepackage{commath}
\makeatletter
\long\def\dddddot#1{%
  {\mathop {#1}\limits ^{\vbox to-1.4\ex@ {\kern -\tw@ \ex@ \hbox {\normalfont .....}\vss }}}%
}
\long\def\multidots#1#2{%
  \count@=0
  {{\mathop {#2}\limits ^{\vbox to-1.4\ex@ {\kern -\tw@ \ex@ \hbox {\normalfont %
  \loop%
  \ifnum#1>\count@%
  .%
  \advance\count@ by1%
  \repeat%
  }\vss }}}}%
}
\makeatother

\begin{document}

\title{\bf Effects of Matter Lagrangian in $f(Q,T)$ Gravity: The Accelerating Cosmological Model}

\author{Rahul Bhagat\orcidlink{0009-0001-9783-9317}}
\email{rahulbhagat0994@gmail.com}
\affiliation{Department of Mathematics, Birla Institute of Technology and Science, Pilani, Hyderabad Campus, Jawahar Nagar, Kapra Mandal, Medchal District, Telangana 500078, India.}
\author{I. V. Fomin \orcidlink{0000-0003-1527-914X}}
\email{ingvor@inbox.ru}
\affiliation{ Bauman Moscow State Technical University, 2-nd Baumanskaya street, 5, Moscow, 105005, Russia}
\author{B. Mishra\orcidlink{0000-0001-5527-3565}}
\email{bivu@hyderabad.bits-pilani.ac.in}
\affiliation{Department of Mathematics, Birla Institute of Technology and Science, Pilani, Hyderabad Campus, Jawahar Nagar, Kapra Mandal, Medchal District, Telangana 500078, India.}

\begin{abstract} We investigate the logarithmic form of $f(Q,T)$ gravity with two different choices of matter Lagrangian such as: $\mathcal{L}_m = p$ and $\mathcal{L}_m = -\rho$.  The parameters of the model has been constrained using Cosmic Chronometers (CC) in combination with DES-SN5YR and Pantheon$^+$ Type Ia supernova datasets. We have observed that the deceleration parameter shows a smooth transition from deceleration to acceleration phase and the effective equation of state parameter ($\omega$) approaches to $-1$ at late times. The $Om(z)$ diagnostic exhibits a decreasing profile, confirming quintessence-like behavior, and the statefinder analysis demonstrates trajectories that remain near the $\Lambda$CDM fixed point but deviate into the quintessence region. The evolution of the density parameters satisfies the flatness condition, and the predicted age of the Universe lies within $t_0 \sim 13.7-14.3$ Gyr, consistent with CMB and stellar estimates.  The findings indicate that the logarithmic model successfully reproduces the late-time accelerated expansion for both the choices of the matter Lagrangian. 

\end{abstract}

\maketitle

\section{Introduction}
Since last two decades, several observational surveys have firmly established that the Universe is currently undergoing accelerated expansion \cite{Spergel_2003_148, Tegmark_2004_69, Hinshaw_2013_208}. The most direct evidence comes from Type Ia supernovae (SNeIa) observations \cite{Riess_1998_116, Perlmutter_1998_517}. Further confirmation arises from Hubble Space Telescope measurements, galaxy redshift surveys, baryon acoustic oscillations, cosmic microwave background and other related probes \cite{Ade_2016_594a, Abbott_2016_460,Larson_2011_192, Komatsu_2011_192}. To explain this late-time acceleration, a variety of theoretical frameworks have been proposed \cite{NOJIRI201159,Sotiriou82451,Ferraro75084031}. In general, these approaches are in two main directions such as introduction of a dark energy(DE) component in the matter sector or the modification of underlying geometry in the gravitational theory \cite{Baudis_2016_43,Bahamonde_2015_92,Capozziello_2008_08}. The basic issue is that the actual nature and origin of DE are still unknown. Also, it is uncertain on the form of DE that best describes the present cosmic behavior. However, according to the cosmological observations the equation of state (EoS) parameter $\omega = -1$ for the $\Lambda$CDM and $\omega < -1$ for the phantom regime. Also, the standard and most successful description so far is the $\Lambda$CDM model \cite{Fujii_1982_26}, which unifies cold dark matter (CDM) with a cosmological constant $\Lambda$ \cite{Yu_2018_856}.  Despite its remarkable agreement with data, $\Lambda$CDM suffers from two major conceptual issues such as the fine-tuning issue and coincidence problems \cite{PhysRevD.101.063531}. The fine-tuning issue refers to the huge discrepancy between the small observed value of the DE density and the larger prediction from quantum field theory \cite{Capozziello_2000_32}. Whereas, the coincidence problem reflects the fact that matter and DE densities happen to be of the same order even though their evolutionary histories are quite different \cite{SCHONEBERG20221}. 

One of the geometrical approaches to modify gravitational theory is to incorporate the nonmetricity, which is the fundamental geometrical quantity\textcolor{red}{~\cite{Nester_1999,Adak:2005cd,Mol:2014ooa,Adak:2018vzk}}. The gravitational theory is termed as symmetric teleparallel gravity. The first modification on the symmetric teleparallel gravity is $f(Q)$ gravity \cite{Jimenez_2018_98}, where the gravitational action be the general function of $Q$. Some of the recent research on $f(Q)$ gravity to address late-time issue are in Ref.\cite{Jimenez_2020_101,Anagnostopoulos_2021_822,Narwade_2024_84,Barros_2020_30,SULTANA2025100422,Narawade_2023_535}. Then with the non-minimal coupling between $Q$ and $T$, the $f(Q,T)$ gravity has been formulated  \cite{Xu_2019_79},  where $T$ represents trace of the energy-momentum tensor of matter. Some of the recent works on $f(Q,T)$ gravity are discussed. 

Shiravand et al. \cite{Shiravand_2024} analyzed the model in the context of early inflation and late-time acceleration using a time-dependent deceleration parameter. Kaczmarek et al. \cite{Kaczmarek_2025} developed its scalar tensor formulation and introduced dynamical variables to investigate the evolution of the Universe. The theory of cosmological perturbations in linear order for  $f(Q,T)$ gravity was considered by Najera and Fajardo in~\cite{Najera:2021afa}.
The implications of matter Lagrangian degeneracy examined \cite{N_jera_2022,Bhagat_2023_42}, whereas Pati et al. \cite{Pati_2023_83} employed dynamical system methods to study the accelerating behavior of cosmological parameters. Teleparallel geometry has been formulated as a gauge theory of translations, sharing the fundamental structure and properties of gauge field theories \cite{Bahamonde_2023_86}. Based on this, to capture the anistropic feature of the Universe, Bianchi-I spacetime in $f(Q,T)$ gravity has been examined in \cite{Zhadyranova_2024_44}. The viability of these models has also been tested against cosmological observations, providing constraints on the model parameters consistent with current data \cite{Rudra_2022}. In the vacuum sector, the theory is shown to admit de Sitter like solutions, with the modified field equations introducing additional correction terms to the standard cosmological dynamics \cite{Khurana_2024,Bhagat_ASPdyna2024}. This gravitational theory has also the ability to reproduce matter bounce scenario inspired by loop quantum cosmology in the FLRW background \cite{Agrawal_2023_83}. Also, the non-singular bouncing scenario of early Universe has been shown in Ref. \cite{Agrawal:2021_bouncing}. In addition, Sharif et al. \cite{Sharif_2024} reconstructed the theory within ghost dark energy scenarios to explore the contribution of dark energy to cosmic dynamics. The versatility of $f(Q,T)$ gravity in addressing key cosmological issues are available in the literature \cite{Zia_2021, Shiravand_2022_37,Bhagat_2023_41,BHAGAT2025101913}. 

In this paper, we intend to investigate the late-time behavior of the Universe in $f(Q,T)$ gravity in two different choices of the matter Lagrangian. To constrain the model parameters, we employ two complementary combinations of observational datasets such as, Cosmic Chronometers (CC) + DES-SN5YR, and CC + Pantheon$^+$. The CC data directly probe the Hubble parameter $H(z)$ in a model-independent manner, while the DES-SN5YR and Pantheon$^+$ supernova compilations serve as standardizable candles mapping the luminosity distance redshift relation. Together, these datasets allow us to place tight bounds on the Hubble constant $H_0$ and other cosmological parameters. The structure of the paper is as follows: In Section \ref{Sec:2}, we present the background equations of $f(Q,T)$ gravity and the field equations are shown for different matter Lagrangian. In Section \ref{Sec:3}, we develop the $f(Q,T)$ gravity model with some well motivated functional form. In Section \ref{Sec:4}, we describe the observational datasets and perform the parameter estimation to obtain the constraints. In Section \ref{Sec:5}, we analyze the dynamical and geometrical parameters of the cosmological models. In Section \ref{Sec:6}, we discuss and provide the conclusions on the models.

\section{Background equations of $f(Q, T)$ theory}\label{Sec:2}

The action of $f(Q,T) $ gravity \cite{Xu_2019_79} is given as,
\begin{eqnarray} \label{Eq.1}
S = \int \left[ \frac{1}{16\pi}f(Q,T)\sqrt{-g}~d^4x + \mathcal{L}_m\sqrt{-g}~d^4x \right],
\end{eqnarray}
where $ \mathcal{L}_m $ represents the matter Lagrangian and $ g $ is the determinant of the metric tensor $ g_{\mu\nu} $. Varying the action \eqref{Eq.1} with respect to the metric tensor leads to the generalized field equations of $f(Q,T)$ gravity,

\begin{multline}\label{Eq.2}
    \frac{2}{\sqrt{-g}} \, \partial_\lambda \left( \sqrt{-g} f_Q P^\lambda_{\mu\nu} \right) - \frac{1}{2} f \, g_{\mu\nu} + f_T (T_{\mu\nu} + \Theta_{\mu\nu}) \\+ f_Q \left( P_{\nu\rho\sigma} Q_\mu^{\rho\sigma} - 2 P^{\rho\sigma\mu} Q_{~\nu}^{\rho\sigma} \right) = \kappa T_{\mu\nu},
\end{multline}

For brevity, we denote the partial derivatives as, $f_Q = \partial f / \partial Q$ and $f_T = \partial f / \partial T$. The superpotential and energy-momentum tensor expressed respectively as,
\begin{eqnarray}\label{Eq.3}
P^{\lambda}_{~\mu\nu} &=& -\frac{1}{2}L^{\lambda}_{~\mu \nu}+\frac{1}{4}\left(Q^{\lambda}-\tilde{Q}^{\lambda}\right)g_{\mu \nu}-\frac{1}{4}\delta^{\lambda}_{(\mu}Q_{\nu)},\nonumber\\
T_{\mu \nu}&=&\frac{-2}{\sqrt{-g}} \frac{\delta(\sqrt{-g}L_{m})}{\delta g^{\mu \nu}},~~~~~~~\Theta_{\mu \nu}=g^{\lambda l}\frac{\delta T_{\lambda l}}{\delta g^{\mu \nu}}.
\end{eqnarray}
and the nonmetricity scalar,
 \begin{eqnarray}\label{Eq.4}
    && Q = Q_{\lambda\mu\nu}P^{\lambda\mu\nu} \nonumber\\
    && =\frac{1}{4}(-Q^{\lambda\nu\rho}Q_{\lambda\nu\rho}+2Q^{\lambda\nu\rho}Q_{\rho\lambda\nu}-2Q^\rho\tilde{Q_\rho}+Q^\rho Q_\rho),
\end{eqnarray}
where $Q_{\lambda} = Q_{\lambda}^{\;\;\mu}\;_{\mu},~ \tilde{Q}_{\lambda}=Q^{\mu}\;_{\lambda\mu}$.\\
Varying the gravitational action \eqref{Eq.1} with respect to the connection yields the field equations of $ f(Q,T) $ gravity as,

\begin{equation}\label{Eq.5}
    \nabla_\mu \nabla_\nu \left( 2\sqrt{-g} \, f_Q \, P^{\mu\nu}_\lambda - \kappa \, H_\lambda^{\mu\nu} \right) = 0,
\end{equation}

where $H^\lambda_{\mu\nu}$ represents the hypermomentum tensor density and can be defined as
\begin{equation}\label{Eq.6}
   H_\lambda^{\mu\nu} = \frac{\sqrt{-g}}{2\kappa} f_T \frac{\delta T}{\delta \Gamma^\lambda_{\mu\nu}} + \frac{\delta(\sqrt{-g} \mathcal{L}_M)}{\delta \Gamma^\lambda_{\mu\nu}}. 
\end{equation}

To investigate how the non-uniqueness in the choice of matter Lagrangian affects the evolution of the cosmic energy density. To this end, we consider the Levi-Civita covariant derivative of the field equations  [Eq.\eqref{Eq.2}]:

\begin{multline}\label{Eq.14}
     D_\mu \big( f_T (T^\mu{}_\nu + \Theta^\mu{}_\nu) - 8\pi T^\mu{}_\nu \big) 
- \tfrac{1}{2} f_T \partial_\nu T
=\\ \frac{1}{\sqrt{-g}} Q_\mu \nabla_\alpha \big(f_Q \sqrt{-g} P^{\alpha\mu}{}_\nu \big)
- \frac{8\pi}{\sqrt{-g}} \nabla_\alpha \nabla_\mu H^{\alpha\mu}{}_\nu ,
\end{multline}

where $\nabla$ denotes the total connection means reducing to the standard derivative in the coincident gauge, while $D_\mu$ refers to the Levi-Civita derivative. Using the connection field Eq. \eqref{Eq.6}, this relation can be rewritten as

\begin{multline}\label{Eq.15}
        D_\mu \big( f_T (T^\mu{}_\nu + \Theta^\mu{}_\nu) - 8\pi T^\mu{}_\nu \big) 
- \tfrac{1}{2} f_T \partial_\nu T
= \\ \frac{1}{\sqrt{-g}} Q_\mu \nabla_\alpha \big(f_Q \sqrt{-g} P^{\alpha\mu}{}_\nu \big)
+ \frac{2}{\sqrt{-g}} \nabla_\alpha \nabla_\mu \big( f_Q \sqrt{-g} P^{\alpha\mu}{}_\nu \big).
\end{multline}

Since the tensor $\Theta^\mu{}_\nu$ appears explicitly in Eq. \eqref{Eq.15}, the matter Lagrangian $L_m$ directly impacts the resulting density evolution. In particular, attention is given to the zeroth component, which determines the time evolution of the energy density, allowing us to examine the implications for the choices of two different matter Lagrangians.\\

We consider an isotropic and homogeneous space time as,

\begin{equation}\label{Eq.7}
ds^2 = -dt^2 + a^2(t)(dx^2+dy^2+dz^2),
\end{equation}

where the scale factor $a(t)$ related to the Hubble parameter $H = \dot{a}/a$. An over dot represents derivative with respect to cosmic time. In this background, the nonmetricity scalar becomes
\begin{equation}\label{Eq.8}
Q = 6H^2 .
\end{equation}

Once the metric and connection are fixed, the generalized Friedmann equations can be derived from the field equations \eqref{Eq.2}, by assuming that the Universe is composed of a perfect fluid. The explicit form of the tensor $T_{\mu\nu}$ and $\Theta_{\mu\nu}$ are respectively 

\begin{eqnarray}\label{Eq.9}
T_{\mu \nu}&=&\frac{-2}{\sqrt{-g}} \frac{\delta(\sqrt{-g}L_{m})}{\delta g^{\mu \nu}}= L_m g_{\mu\nu} - 2\frac{\delta L_m}{\delta g^{\mu\nu}},\nonumber\\
    \Theta_{\mu\nu} &=& g^{\alpha\beta}\frac{\delta T_{\alpha\beta}}{\delta g^{\mu\nu}} 
= L_m g_{\mu\nu} - 2T_{\mu\nu}.
\end{eqnarray}

One can see that the structure of $T_{\mu\nu}$ and $\Theta_{\mu\nu}$ depends directly on the choice of the matter Lagrangian $L_m$. For a perfect fluid stress energy tensor \eqref{Eq.3}, there are two commonly used forms of $L_m$ such as $L_m = p$,  the pressure or $L_m = -\rho$, the energy density. This non-uniqueness in $L_m$ propagates into the field equations, leading to different forms of the generalized Friedmann equations.\\

\subsection*{Case I: $L_m = p$} 

For $L_m = p$, the tensor becomes $\Theta_{\mu\nu} = p g_{\mu\nu} - 2T_{\mu\nu}.$
The resulting generalized Friedmann equations are

\begin{equation}\label{Eq.10}
     f - 6 f_Q H^2 = 8\pi \rho + f_T(\rho + p),    
\end{equation}

\begin{equation}\label{Eq.11}
      -8\pi p + 2(H \dot{f}_Q + \dot{f}_Q H) = f - 6 f_Q H^2.     
\end{equation}

Adopting the coincident gauge and setting $L_m = p$, the zeroth component of Eq. \eqref{Eq.15} yields

\begin{equation}\label{Eq.16}
    \dot{\rho} = 
-\frac{3H (f_T + 8\pi) \rho (1+  p/\rho)}{8\pi + \tfrac{1}{2} f_T (3 - \dot{p}/\dot{\rho})- f_{TT} \rho (1+ p/\rho)(1-3~\dot{p}/\dot{\rho})} 
\end{equation}

where $f_{TT} \equiv d^2 f/dT^2 = d f_T/dT$, when $f_T = 0$, this reduces to the standard continuity equation:

\begin{equation}\label{Eq.17}
    \dot{\rho} + 3H \rho (1+w) = 0.
\end{equation}

 In the general case, the coupling to $T$ introduces extra contributions, which may be interpreted as an exchange of energy between matter and geometry.

\subsection*{Case II: $L_m = -\rho$}

For $L_m = -\rho$, the tensor reads $\Theta_{\mu\nu} = -\rho g_{\mu\nu} - 2T_{\mu\nu}.$
The corresponding Friedmann equations become

\begin{equation}\label{Eq.12}
   f - 6 f_Q H^2 = 8\pi \rho,    
\end{equation}

\begin{equation}\label{Eq.13}
    f - 2(\dot{f}_Q H + f_Q(\dot{H} + 3H^2)) = -8\pi p - f_T(\rho+p). 
\end{equation}

In the above cases, the first Friedmann equation \eqref{Eq.12} is differ to that of $f(Q)$ gravity, showing explicit matter geometry coupling in contrast to the previous case.

For $L_m = -\rho$, the zeroth component of Eq. \eqref{Eq.15} leads instead to

\begin{equation}\label{Eq.18}
    \dot{\rho} = 
-\frac{3H (f_T + 8\pi) \rho (1+ p/\rho)}{8\pi + \tfrac{1}{2} f_T (1 - 3~\dot{p}/\dot{\rho})} .
\end{equation}

In this formulation, the second derivative term $f_{TT}$ does not appear, which simplifies the structure of the continuity equation. Consequently, solving the density evolution equation is technically more straightforward in this case compared to the $L_m = p$ scenario.

\section{The $f(Q,T)$ gravity model} \label{Sec:3}
To investigate the cosmological implications with different forms of matter Lagrangian, we consider the functional form of $f(Q,T)$ \cite{Najera_2023_524} as
\begin{equation}\label{Eq.19}
f(Q,T) = -Q - \alpha Q_0 \ln\left(\frac{Q}{Q_0}\right) + 16\pi \beta T ,
\end{equation}

where $\alpha$, $\beta$ are model parameters and $Q_0=6 H_0^2$, where $H_0$ is a present rate of expansion of the Universe. When $\alpha = \beta = 0$, one can recover general relativity. The nonzero values of $\alpha$ and $\beta$ can be interpreted as allowing an exchange of energy between matter and geometry. Now, to frame the cosmological model using numerical methods, we need to assume some appropriate sets of dimensionless and model-independent variables as,

\begin{equation}\label{Eq.20}
h=\frac{H}{H_0},\ \ \, x=\frac{\rho_m}{3H^2}.
\end{equation}

Before we incorporate the dimensionless variables, it is worthwhile to mention that the field equations Eq. \eqref{Eq.10} and Eq. \eqref{Eq.11} can also be expressed in redshift using the transfomation, 

\begin{equation}\label{Eq.30}
H(z) = -\frac{1}{1+z}\frac{dz}{dt}.
\end{equation}

In observational cosmology, this relation forms the basis of the cosmic chronometer method, where the derivative $dt/dz$ is inferred from age-dating of passively evolving galaxies. However, in our analysis, Eq. (\ref{Eq.30}) is only employed as a mathematical transformation to express the system in terms of redshift rather than time. 

For $L_m = p$. In a dust-dominated Universe, using Eq. \eqref{Eq.30} and Eq. \eqref{Eq.20} in the field equations [\eqref{Eq.10}-\eqref{Eq.11}], one can write the resulting dynamical equations as,

\begin{equation}\label{Eq.21}
\frac{dh(z)_{L_m=p}}{dz}=\frac{3 (2 \beta +1) h(z)^3 x(z)}{2 (z+1) \left(\alpha +h(z)^2\right)},
\end{equation}
and
\begin{equation}\label{Eq.22}
\frac{dx(z)_{L_m=p}}{dz}= \frac{3 x(z)}{z+1} \left(\frac{2 \beta +1}{3 \beta +1}-\frac{(2 \beta +1) h(z)^2 x(z)}{\alpha +h(z)^2}\right)
\end{equation}

Also, the deceleration parameter ($q(z)$) and EoS parameter ($\omega(z)$) can be obtained in terms of  dimensionless variable as, 

\begin{equation}\label{Eq.26}
q(z)_{L_m=p}=\frac{3 (2 \beta +1) h(z)^2 x(z)}{2 \left( \alpha +h(z)^2\right)}-1
\end{equation}
and 
\begin{equation}\label{Eq.27}
\omega(z)_{L_m=p}=\frac{(2 \beta +1) h(z)^2 x(z)}{ \alpha +h(z)^2}-1
\end{equation}

Similarly, for $L_m = -\rho$, using Eq. \eqref{Eq.30} and Eq. \eqref{Eq.20} in the field equations [\eqref{Eq.12}-\eqref{Eq.13}], the resulting dynamical equations become,

\begin{equation}\label{Eq.23}
\frac{dh(z)_{L_m=-\rho}}{dz}=\frac{3 (2 \beta +1) h(z)^3 x(z)}{2 (z+1) \left(2 \alpha +h(z)^2\right)},
\end{equation}
and
\begin{equation}\label{Eq.24}
\frac{dx(z)_{L_m=-\rho}}{dz}= \frac{3 x(z)}{z+1} \left(\frac{2 \beta +1}{\beta +1}-\frac{(2 \beta +1) h(z)^2 x(z)}{2 \alpha +h(z)^2}\right)
\end{equation}

Subsequently,

\begin{equation}\label{Eq.28}
    q(z)_{L_m=-\rho}=\frac{3 (2 \beta +1) h(z)^2 x(z)}{2 \left(2 \alpha +h(z)^2\right)}-1
\end{equation}
and 
\begin{equation}\label{Eq.29}
   \omega(z)_{L_m=-\rho}=\frac{(2 \beta +1) h(z)^2 x(z)}{2 \alpha +h(z)^2}-1
\end{equation}

These expressions allow us to directly examine the transition from decelerated to accelerated expansion and the dynamical nature of the effective dark energy component in the two formulations of the matter Lagrangian.

Now, we shall solve the coupled differential system of equations numerically with the initial conditions $h_0 = 1$ and $x_0 = \Omega_{m0}$. Within the framework of the logarithmic $f(Q,T)$ model \eqref{Eq.19}, this procedure introduces four independent free parameters such as, the present-day Hubble constant $H_0$, the matter density parameter $\Omega_{m0}$, and the model parameters $\alpha$ and $\beta$. Once the system of equations is solved for both matter Lagrangian choices, the evolution of the dimensionless density parameters can be determined as
\begin{equation}\label{Eq.25}
\Omega_m(z) = x(z), \quad \Omega_f(z) = 1 - x(z),
\end{equation}

where $\Omega_m(z)$ represents the contribution of matter and $\Omega_f(z)$ corresponds to the effective dark energy sector induced by modified gravity effects.

\section{The Observational Analysis}\label{Sec:4}
We shall investigate the cosmological implications of the models in the Bayesian statistical framework that uses the Markov Chain Monte Carlo (MCMC) method. The analysis would be carried out with the \texttt{emcee} Python package \cite{Foreman-Mackey_2013_125}. This will efficiently sample the posterior distributions of the model parameters. The numerical method would constrain the free parameters of the model directly from the datasets. The datasets to be used are cosmic chronometers (CC), DES-SN5YR, and Pantheon$^+$ supernovae. The MCMC algorithm generates chains where each step depends on the likelihood of the previous one, thereby systematically exploring the multidimensional parameter space. Using this approach, one can observe that one-dimensional posterior distributions identify the most probable ranges of individual parameters, whereas the two-dimensional confidence contours illustrate correlations and possible degeneracies. The final bounds of the model parameters are reported with $1\sigma$ and $2\sigma$ confidence interval. We provide here a brief description of each of the cosmological datasets.\\

The {\bf CC technique} provides direct, model-independent probe of the Hubble expansion rate $H(z)$. We employ 32 CC measurements compiled in \cite{Moresco_2022_25}, with the likelihood quantified via the chi-squared estimator
\begin{equation}\label{Eq.31}
\chi^2_{\text{CC}} = \sum_{i=1}^{32} \frac{\big(H(z_i,\Theta) - H_{\text{obs}}(z_i)\big)^2}{\sigma_H^2(z_i)},
\end{equation}
where $H(z_i,\Theta)$ is the theoretical value predicted by the set of parameters $\Theta = \{H_0, \Omega_{m0}, \alpha, \beta\}$. The $H_{\text{obs}}(z_i)$ are the observed Hubble values with uncertainties $\sigma_H(z_i)$. \\

The {\bf Pantheon$^+$} compilation \cite{Brout_2022_938} is with 1701 Type Ia supernovae data points that spans over the redshift range $0.01 < z < 2.26$.  Whereas the {\bf Dark Energy Survey supernovae 5-Year} (DES-SN5YR) \cite{DEScollaboration2025} sample complements with 1829 Type Ia supernova events in the range $0.025 \leq z \leq 1.13$. This includes 1635 discovered by DES and 194 from external low-redshift surveys. This dataset is especially effective at probing the intermediate redshifts. For both Pantheon$^+$ and DES-SN5YR the relevant observable is the distance modulus,
\begin{equation}\label{Eq.32}
\mu(z_i,\Theta) = 5 \log_{10}[D_L(z_i,\Theta)] + 25,
\end{equation}
in which the luminosity distance is
\begin{equation}\label{Eq.33}
D_L(z_i,\Theta) = c(1+z_i)\int_0^{z_i}\frac{dz'}{H(z',\Theta)}.
\end{equation}

The respective Chi-squared function becomes,

\begin{equation}\label{Eq.34}
\chi^2_{\text{Pantheon$^+$}} = (\Delta \mu(z_i,\Theta))^T C_{\text{Pantheon$^+$}}^{-1} (\Delta \mu(z_i,\Theta)),
\end{equation}
and
\begin{equation}\label{Eq.35}
\chi^2_{\text{DES}} = (\Delta \mu(z_i,\Theta))^T C_{\text{DES}}^{-1} (\Delta \mu(z_i,\Theta)),
\end{equation}
where $\Delta \mu(z_i,\Theta)$ denotes the residuals between theoretical predictions and observed values, and $C^{-1}$ encodes both statistical and systematic errors. To strengthen the constraint, we perform the analysis by pairing datasets such as CC + Pantheon$^+$ and CC + DES-SN5YR. Now, the total chi-squared becomes

\begin{equation}\label{Eq.36}
\chi^2_{\text{tot}} = \chi^2_{\text{CC}} + \chi^2_{\text{Pantheon$^+$}} \quad \text{or} \quad \chi^2_{\text{tot}} = \chi^2_{\text{CC}} + \chi^2_{\text{DES}},
\end{equation}

This combined likelihood analysis may provide tight constraints on the parameter set $\Theta$, specifically the Hubble constant $H_0$, the present matter density $\Omega_{m0}$, and the model parameters $\alpha$ and $\beta$. By unifying independent observational probes, we not only reduce parameter degeneracies but also test the internal consistency of the model across different redshift regimes.\\

\subsection{The Results:} 
Using this procedure, we obtain the contour plot (FIG. \ref{Fig1}) of both combinations of datasets for two different matter Lagrangian. The constrained values of the free parameters are provided in TABLE--\ref{tableA1}. The present value of the Hubble parameter for all the cases are in the range of ($68.77^{+0.66}_{-0.67}$, $71.91\pm 0.53$) and the density parameter for matter ranges from ($0.300\pm 0.023, 0.318^{+0.051}_{-0.070}$). The whisker plot (FIG.\ref{fig02}) further shows the variation on the present value of these cosmological and model parameters for different matter Lagrangian.  In FIG. \ref{Fig21}, for $L_m=p$, we have presented the error bar plots for the datasets $H(z)$ and the distance modulus plots for the datasets Pantheon$^+$ and DES-SN5YR. The curves of the models obtained and compared with the $\Lambda$CDM curve. It has been observed that all the plots are traversing within the error bars. Both the models curves are aligning closely with minimal deviation between them in the CC data whereas it fits excellently across the redshift in Pantheon$+$ and DES-5NYR datasets. Similar analysis has been formed for the matter Lagrangian $L_m=-\rho$ and shown in  FIG.\ref{Fig3}. 

\begin{widetext}
~\begin{figure}[H]
\centering
\includegraphics[width=17cm,height=15cm]{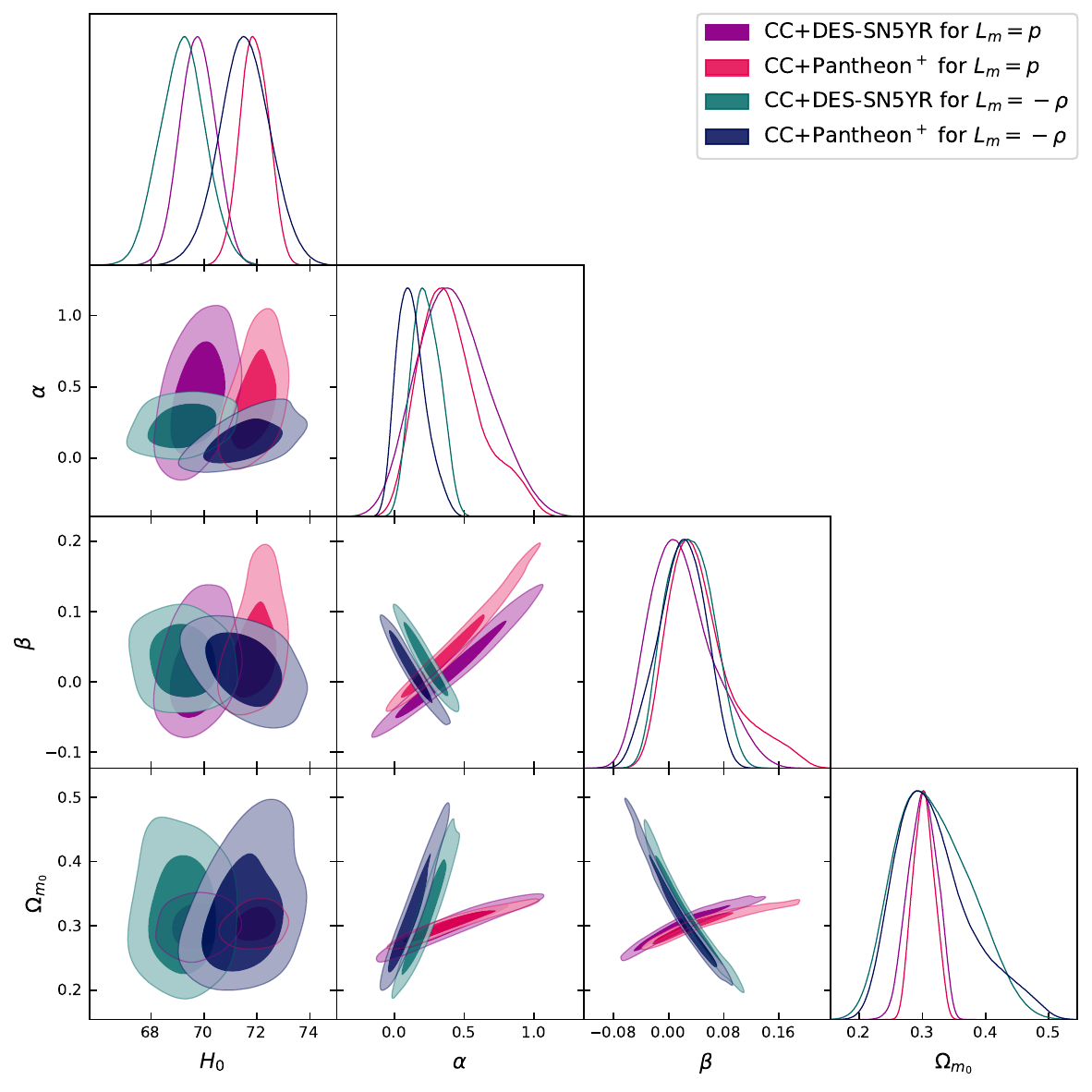}
\caption{The contour plots display $1\sigma$ and $2\sigma$ uncertainty regions for the model parameters $H_0$, $\alpha$, $\beta$ and $\Omega_{m_0}$. The contours are based on the combined sample of  CC+DES5YR and CC+Pantheon$^+$.}
\label{Fig1}
\end{figure}

\begin{table*}[htb]
\renewcommand\arraystretch{1.5}
\centering % used for centering table
{
\begin{tabular}{|c|c|c|c|c|} % centered columns (3 columns)
\hline %inserts double horizontal lines
~~~Observational Datasets~~~& ~~~$H_0$  ~~~& ~~~~$\alpha$~~~~ & ~~~$\beta$~~~& ~~~~$\Omega_{m_0}$~~~~
     \\ [0.5ex] % inserts table
%heading
% inserts single horizontal line
\hline\hline
CC+DES-SN5YR for $L_m = p$  & $68.77^{+0.66}_{-0.67}$ &  $0.42^{+0.23}_{-0.29}$ & $0.019^{+0.034}_{-0.055}$&$0.300\pm{0.023}$ \\
\hline
 CC+Pantheon$^+$ for $L_m = p$ & $71.91\pm0.53$ &  $0.40^{+0.15}_{-0.27}$ & $0.046^{+0.027}_{-0.057 }$&$0.302^{+0.017}_{-0.018}$ \\
\hline
CC+DES-SN5YR for $L_m = -\rho$ & $69.20\pm0.85$ &  $0.225^{+0.098}_{-0.110}$ & $0.030\pm0.033$&$0.318^{+0.051}_{-0.070}$ \\
\hline
CC+Pantheon$^+$ for $L_m = -\rho$ & $71.54^{+0.95}_{-0.98}$ &  $0.121^{+0.080}_{-0.120}$ & $0.019^{+0.038}_{-0.031}$& $0.309^{+0.039}_{-0.075}$\\
 \hline %inserts single line
\end{tabular}}
\caption{Summary of MCMC constrained cosmological parameters from CC+DES-SN5YR and CC+Pantheon$^+$ datasets. The table contains the posterior estimates of the Hubble constant ($H_0$), matter density parameter ($\Omega_{m_0}$), and the model parameters $\alpha$, $\beta$ with their $1\sigma$ confidence intervals.} 

\label{tableA1}
\end{table*}

\begin{figure}[htbp]
\centering
\includegraphics[width=1.1\textwidth]{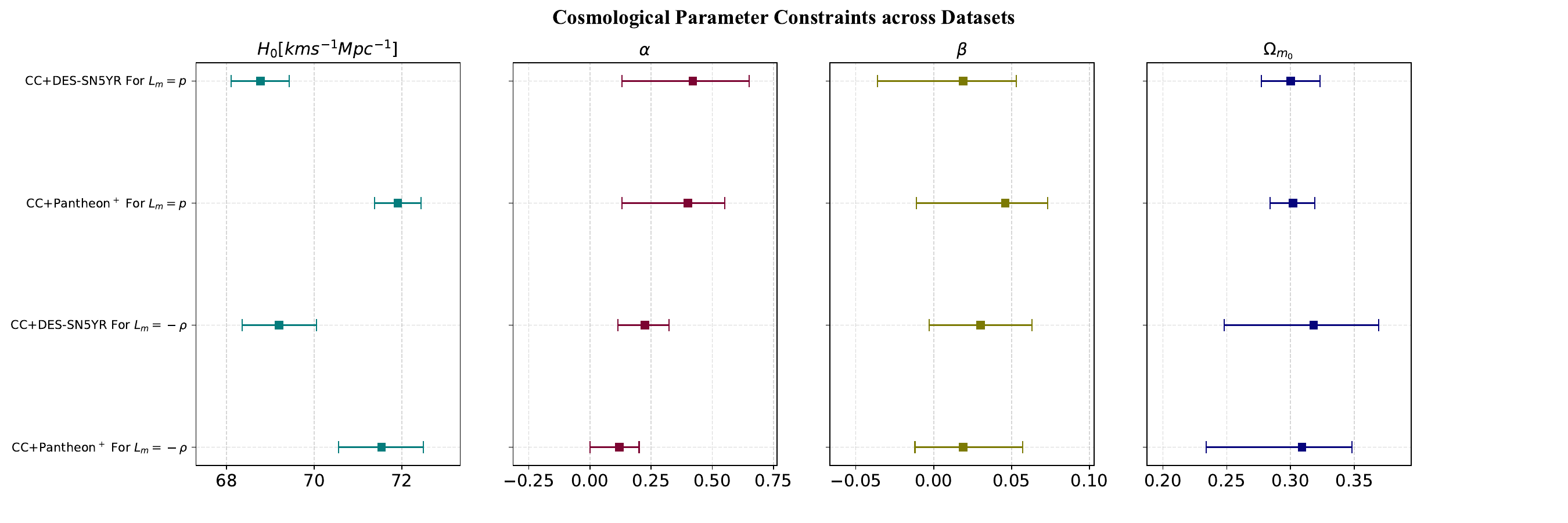}
\caption{Whisker plot showing the constraints on cosmological parameters for different matter Lagrangian.}
\label{fig02}
\end{figure}
\end{widetext}

\begin{widetext}

\begin{figure}[H]
\centering
\includegraphics[width=10cm,height=6cm]{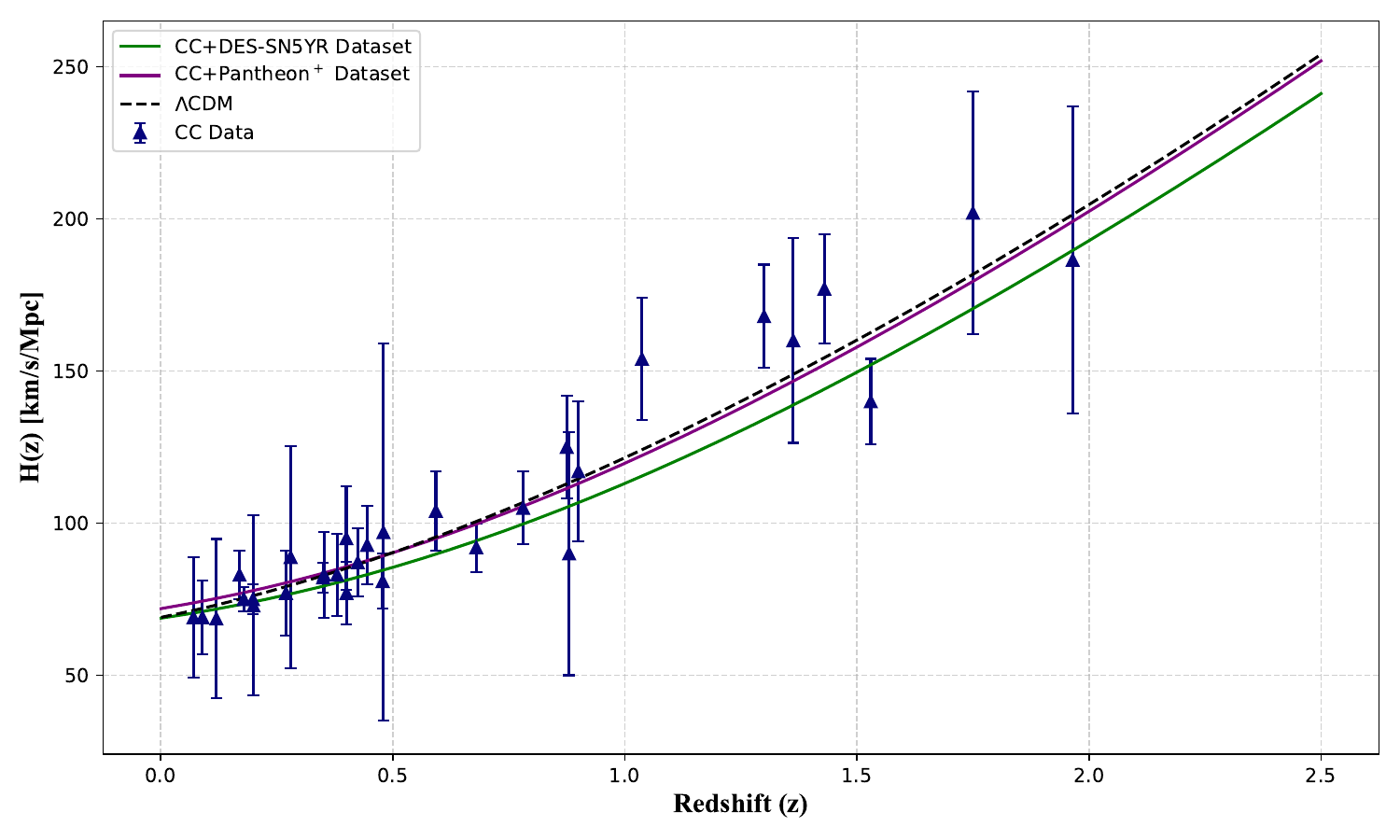}
\includegraphics[width=8cm,height=5cm]{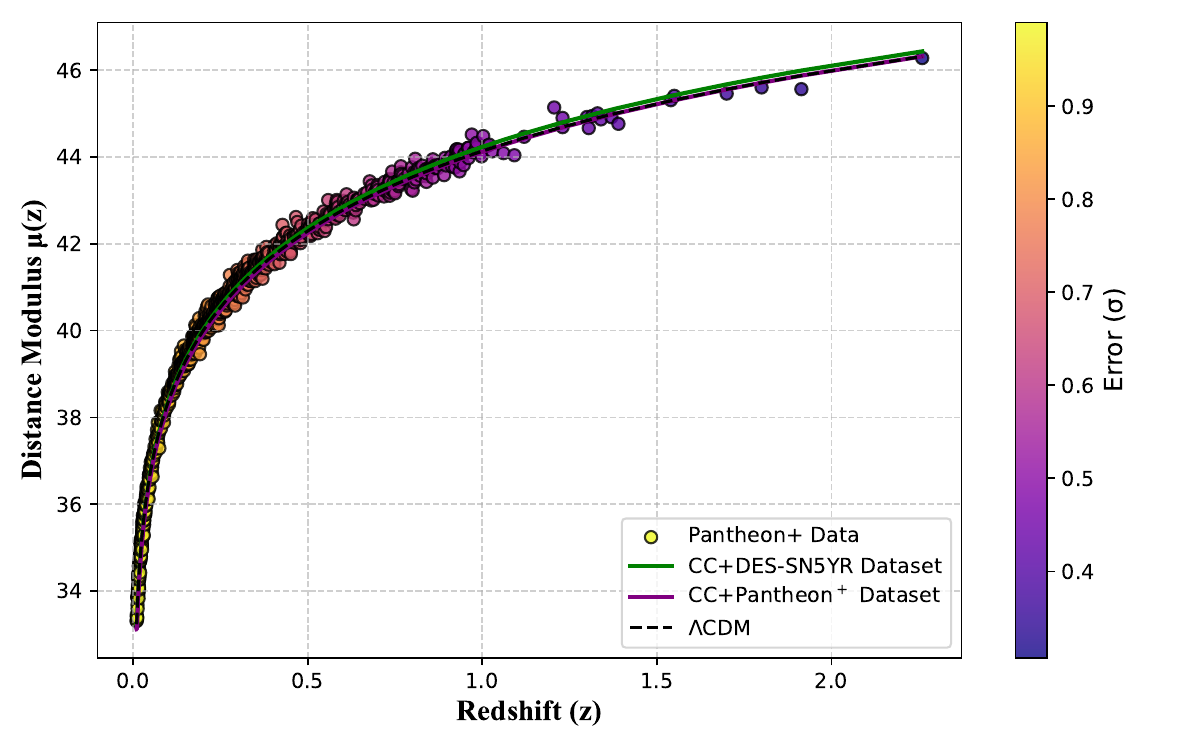}
\includegraphics[width=8cm,height=5cm]{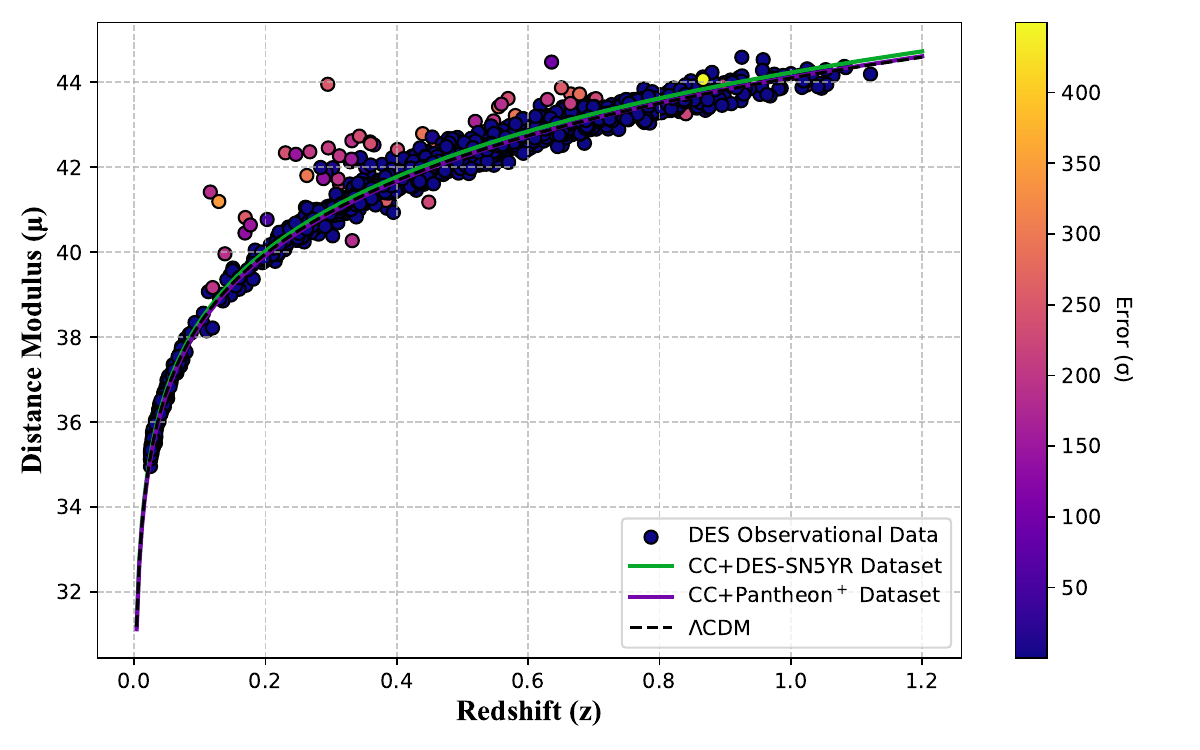}
\caption{{\bf For $L_m=p$}: {\bf Upper Panel:} The error bars show the uncertainty for 32 data points from CC sample. {\bf Left Panel:} Distance modulus from Pantheon$^+$ dataset. {\bf Right Panel:} Distance modulus from DES-SN5YR dataset. The violet curve represents the model obtained through CC+Pantheon$^+$ dataset. The green curve represents the model obtained through CC+DES-SN5YR dataset. The broken black line represents $\Lambda$CDM.}
\label{Fig21}
\end{figure}

\end{widetext}
\begin{widetext}

   \begin{figure}[H]
\centering
\includegraphics[width=10cm,height=6cm]{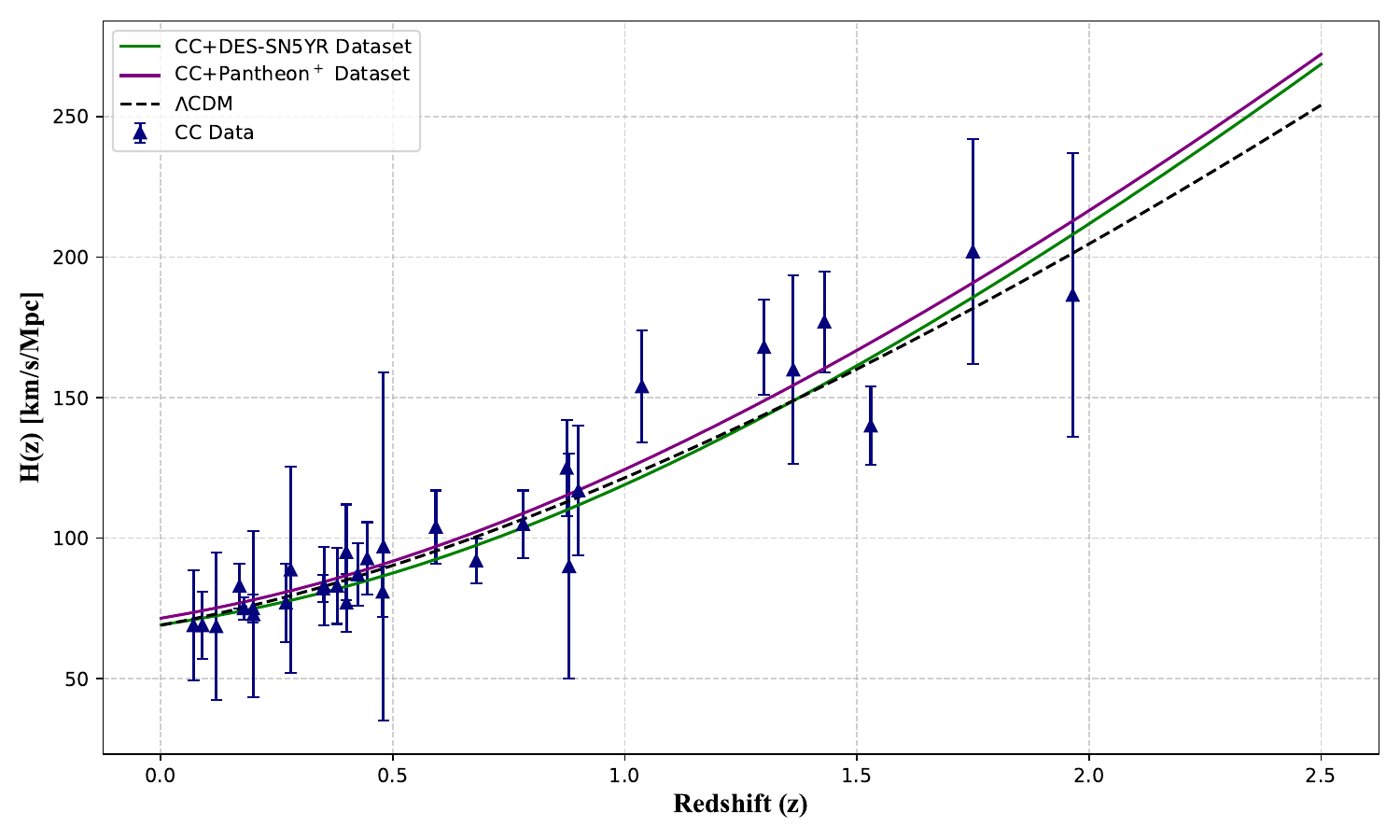}
\includegraphics[width=8cm,height=5cm]{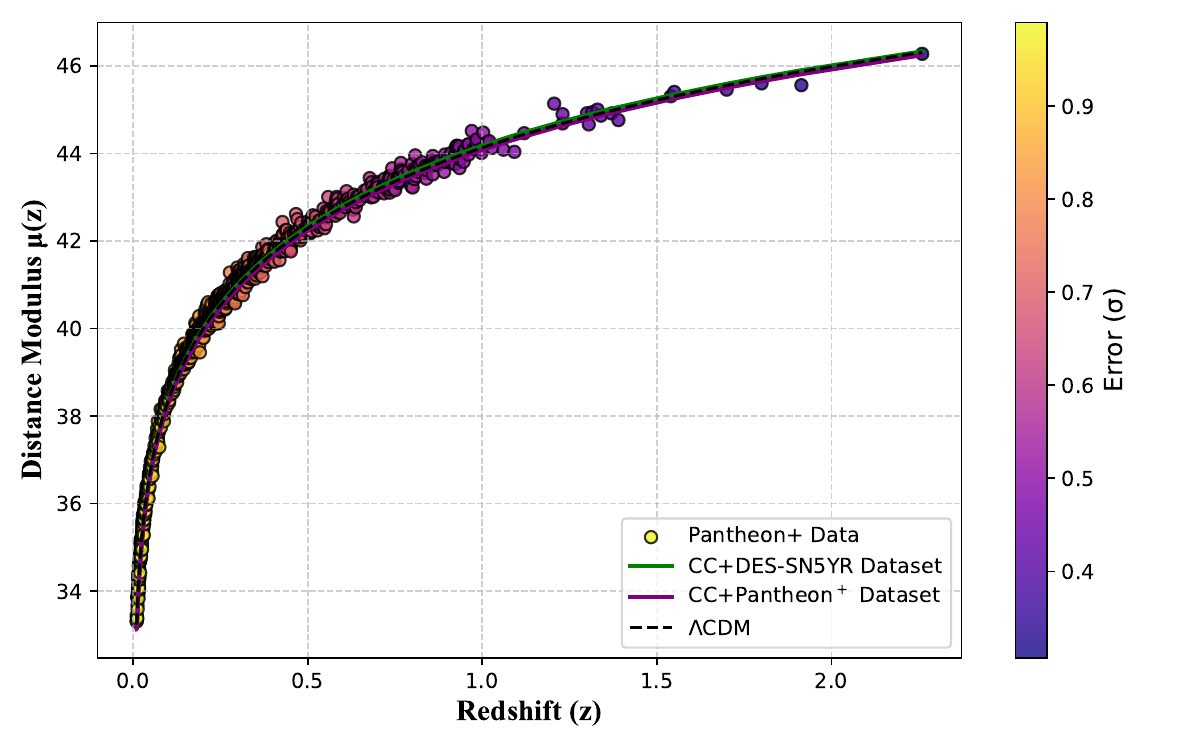}
\includegraphics[width=8cm,height=5cm]{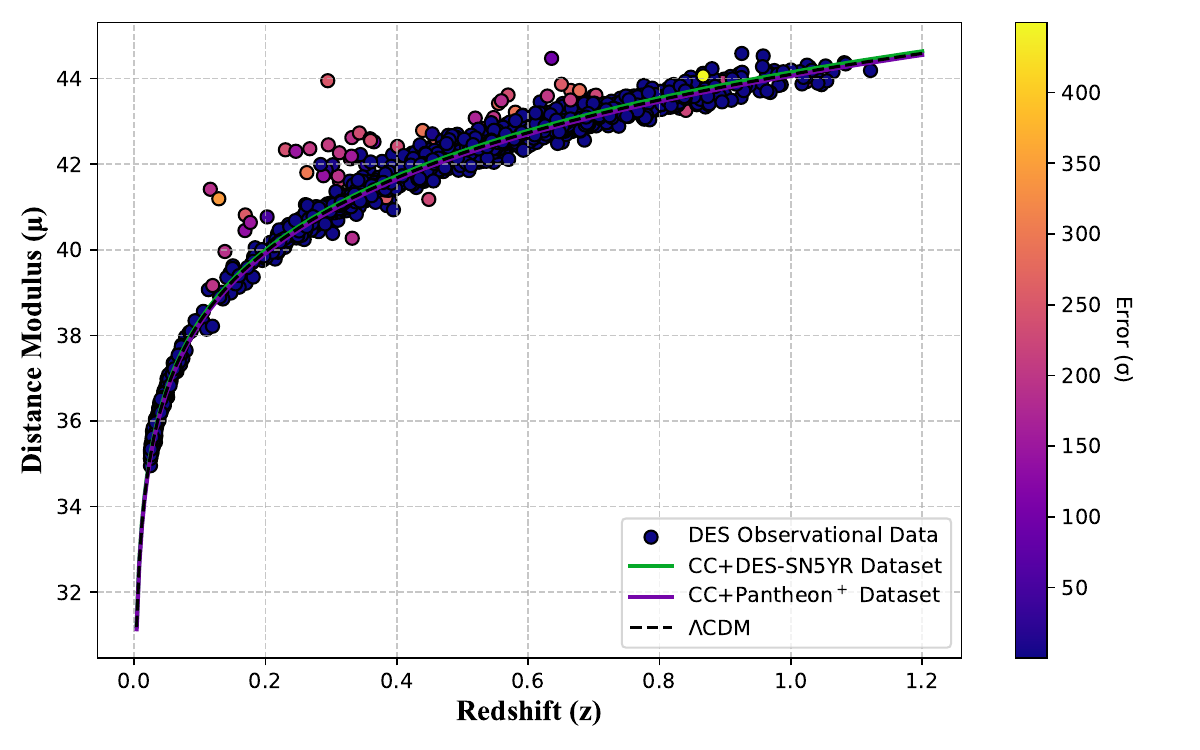}
\caption{{\bf For $L_m=-\rho$}: {\bf Upper Panel:} The error bars show the uncertainty for 32 data points from CC sample; {\bf Left Panel:} Distance modulus from Pantheon$^+$ dataset; {\bf Right Panel:} Distance modulus from DES-SN5YR dataset. The violet curve represents the model obtained through CC+Pantheon$^+$ dataset. The green curve represents the model obtained through CC+DES-SN5YR dataset. The broken black line represents $\Lambda$CDM.}
\label{Fig3}
\end{figure}
\end{widetext}

\section{Cosmological Parameters}\label{Sec:5}

 The present value of the cosmological parameters as observed from the observational analysis is listed in TABLE--\ref{table:2} for different combinations of the datasets. The transient evolutionary behavior of the deceleration parameter ($q(z)$) is shown in FIG. \ref{Fig4}). The curve shows from early deceleration to late time acceleration and the transition occurs at different points but in a narrow range for different combinations of datasets. However, at a late time all curves converge to $-1$. The evolutionary behavior remains similar regardless of the chosen matter Lagrangian, except for the fact that the evolution starts from a different $q$-value. For $\mathcal{L}_m = p$, we find $q_0 = -0.671$ with $z_{\text{tr}} = 0.763$, while another branch yields $q_0 =-0.646$ with the $z_{\text{tr}} = 0.753$ transition redshift. For $ \mathcal{L}_m = -\rho$, the results are $q_0 = -0.651$ with $z_{\text{tr}} = 0.647$, and $q_0 = -0.612$ with $z_{\text{tr}} = 0.646$. In FIG. \ref{Fig5}, it can be observed that for both the matter Lagrangian, the EoS parameter $\omega$ approaches $\ -1$, indicating the dominance of a dark energy epoch at late times. The present values of the EoS parameter are $\omega_0 = -0.780$ (CC+DES-SN5YR) and $\omega_0 = -0.764$ (CC+Pantheon$^+$) for the case $L_m = p$, while for $L_m = -\rho$ the corresponding values are $\omega_0 = -0.767$ (CC+DES-SN5YR) and $\omega_0 = -0.741$ (CC+Pantheon$^+$), respectively.   
\begin{widetext}
 \begin{table}[htb]
\renewcommand\arraystretch{1.5}
\centering % used for centering table
\begin{tabular}{|c|c|c|c|c|}
\hline
Cosmological Parameter & \parbox[c][1cm]{2.7cm}{CC+DES-SN5YR for $L_m = p$} & \parbox[c][1cm]{2.7cm}{ CC+Pantheon$^+$ for $L_m = p$}& \parbox[c][1cm]{2.7cm}{CC+DES-SN5YR for $L_m = -\rho$} &  \parbox[c][1cm]{2.7cm}{CC+Pantheon$^+$ for $L_m =  -\rho$}\\
\hline
$q_0$  & $-0.671$ & $-0.646$ & $-0.651$ & $-0.612$ \\
\hline
$z_{\mathrm{tr}}$ & $0.763$ & $0.753$ & $0.647$ & $0.646$ \\
\hline
$w_0$  & $-0.780$ & $-0.764$ & $-0.767$ & $-0.741$ \\
\hline
$s_0$  & $-0.013$ & $-0.008$ & $-0.030$ & $-0.024$ \\
\hline
$r_0$  & $1.046$ & $1.028$ & $1.105$ & $1.079$ \\
\hline
Age of the Universe [Gyr] & $14.383$ & $13.917$ & $13.800$ & $13.731$ \\
\hline
\end{tabular}
\caption{Present values of the cosmological parameters.}
\label{table:2}
\end{table}
\end{widetext}
 
\begin{widetext}
~\begin{figure}[H]
\centering
\includegraphics[width=18cm]{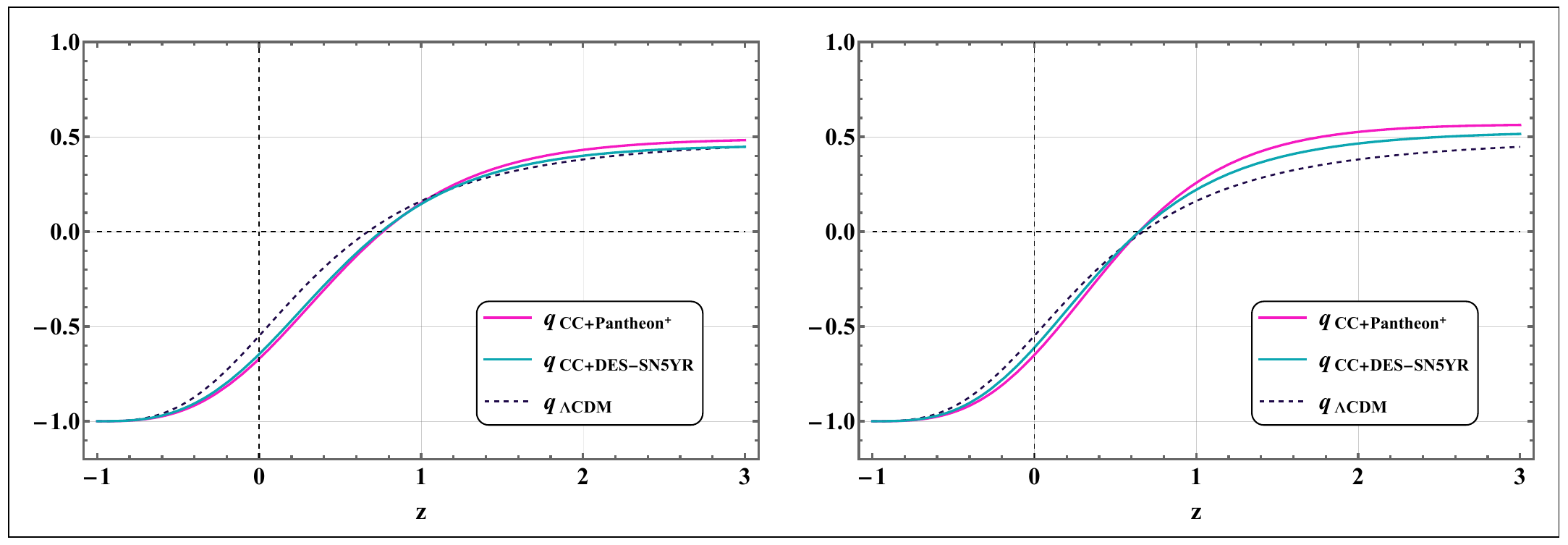}
\caption{Evolution of the deceleration parameter $q(z)$ in redshift. {\bf Left Panel:} $L_m = p$, {\bf Right Panel:} $L_m = -\rho$.}
\label{Fig4} 
\end{figure}
\end{widetext}

\begin{widetext}
      ~\begin{figure}[H]
    \centering
    \includegraphics[width=18cm]{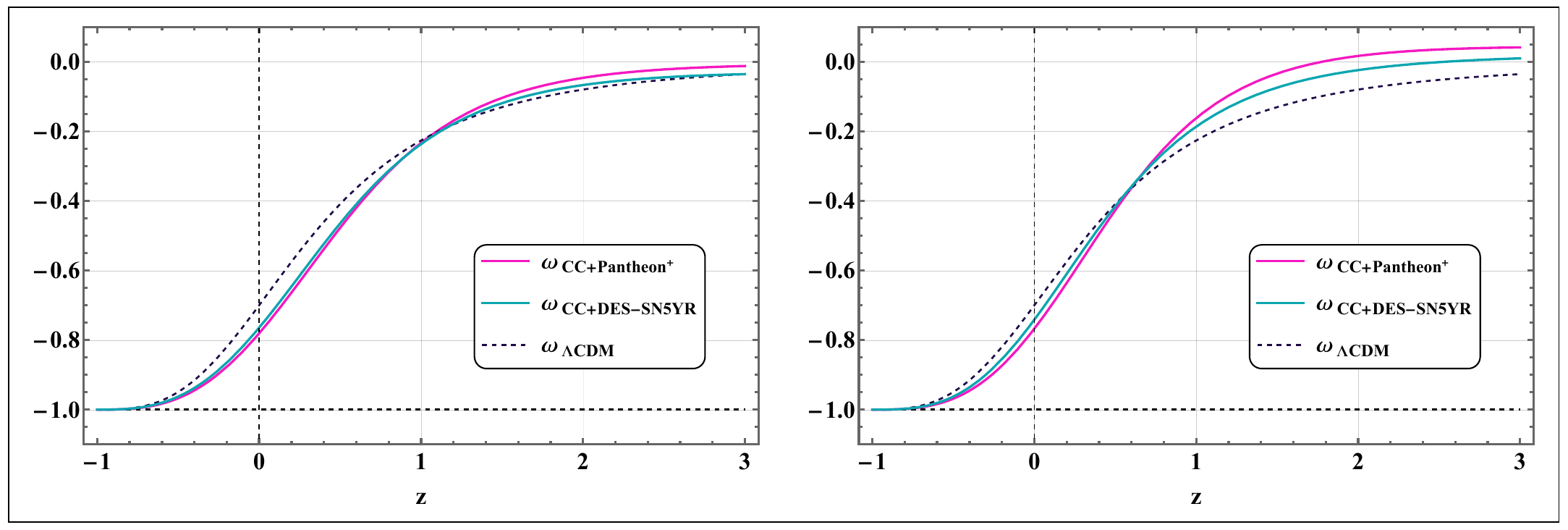}
    \caption{Evolution of the EoS parameter $\omega(z)$ in redshift. {\bf Left Panel:} $L_m = p$, {\bf Right Panel:} $L_m = -\rho$.}
    \label{Fig5} 
 \end{figure}
\end{widetext}

Fig. \ref{Fig6} illustrates the evolution of the density parameters of the models under the two matter Lagrangian choices, $\mathcal{L}_m = p$ (Left Panel) and $\mathcal{L}_m = -\rho$ (Right Panel), using the combined  CC+DES-SN5YR and CC+Pantheon$^+$ datasets. In both cases, we plot the dark energy density parameter $\Omega_f$ and the matter density parameter $\Omega_m$, and compare them with the standard $\Lambda$CDM results ($\Omega_\Lambda$ and $\Omega_m$). For both datasets, the matter density $\Omega_m$ decreases monotonically with redshift, while the effective contribution $\Omega_f$ grows at late times, approaching unity, thereby reproducing the observed accelerated expansion. The transition behavior is consistent across both matter Lagrangian prescriptions, although small quantitative differences arise: in the case of $\mathcal{L}_m = -\rho$, the growth of $\Omega_f$ is slightly faster, leading to a stronger dark energy dominance at low redshift, while for $\mathcal{L}_m = p$, the evolution remains closer to the $\Lambda$CDM trajectory. Importantly, in both scenarios the total contribution satisfies the closure condition $\Omega_f + \Omega_m \approx 1$, consistent with a spatially flat Universe as supported by observations. The agreement with GR at high redshift and the deviation towards a dark energy dominated phase at low redshift indicate that the models can effectively mimic $\Lambda$CDM while allowing additional flexibility in fitting different observational datasets.

\begin{widetext}
      ~\begin{figure}[H]
    \centering
    \includegraphics[width=18cm]{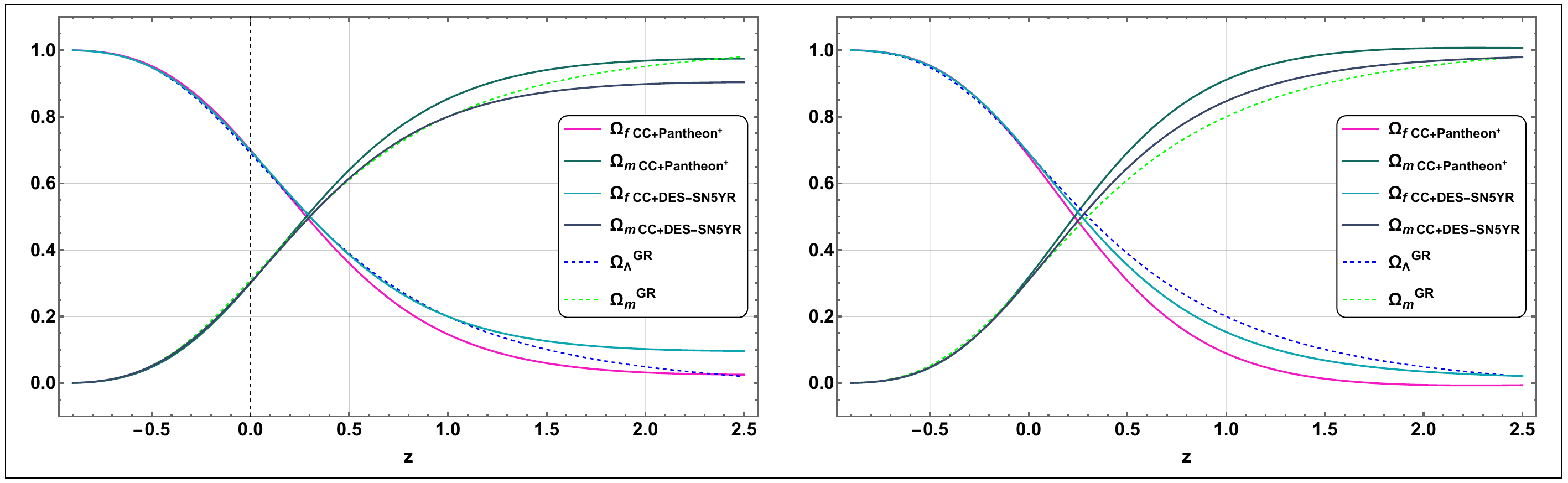}
    \caption{Evolution of matter density parameter $\Omega_m(z)$ and dark energy density parameter $\Omega_f(z)$ in redshift. {\bf Left Panel:} $L_m = p$, {\bf Right Panel:} $L_m = -\rho$.}
    \label{Fig6} 
 \end{figure}
\end{widetext}

The $Om(z)$ diagnostic is an important consistency check for cosmological models, which can be expressed as, $Om(z)=\frac{h^2(z) - 1}{(1+z)^3 - 1}$. It serves as a null test for the $\Lambda$CDM scenario and in the standard model, $Om(z)$ remains constant and corresponds directly to present matter density parameter $\Omega_{m0}$. Deviations from constancy therefore indicate departures from $\Lambda$CDM, with an increasing trend signaling phantom-like behavior and a decreasing trend pointing to quintessence-like dynamics. In Fig. \ref{Fig7}, it can be seen that the $Om(z)$ function exhibits a clear decreasing profile with redshift, demonstrating that the effective dark energy behaves as quintessence. This result highlights that, while the model successfully follows the background expansion history of $\Lambda$CDM, it also allows for richer dynamical features, with a time-varying effective dark energy EoS consistent with a quintessence-like scenario.

\begin{widetext}
     ~\begin{figure}[H]
    \centering
    \includegraphics[width=18cm]{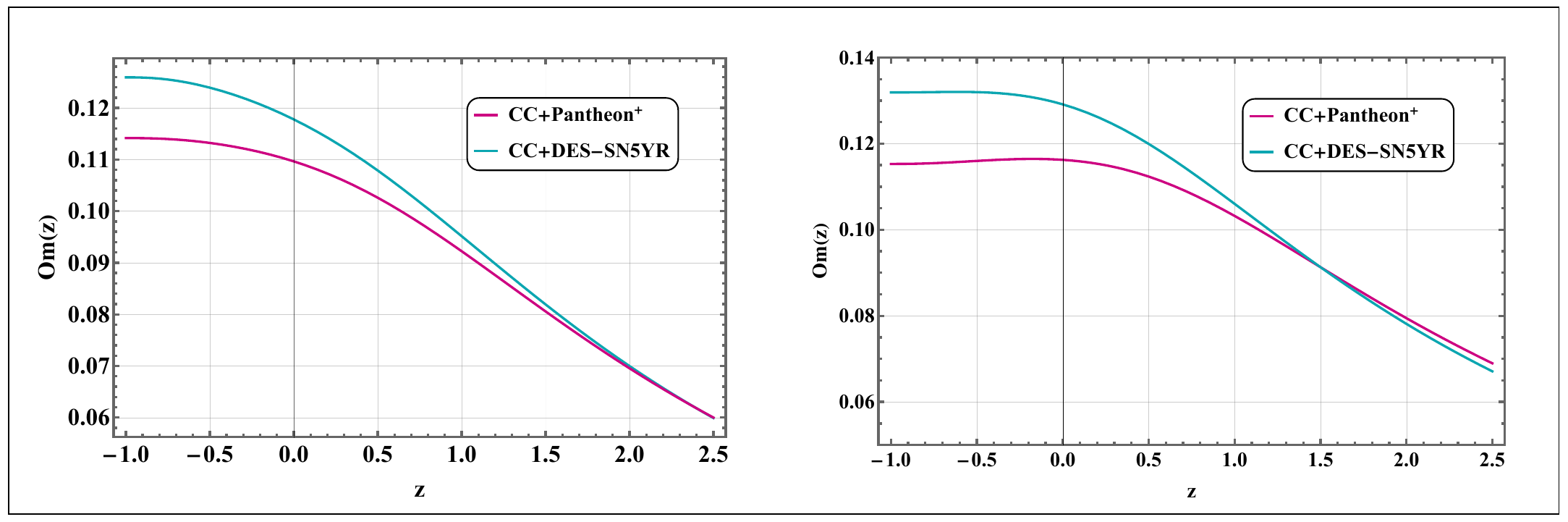}
    \caption{Evolution of the $Om(z)$ diagnostic in redshift {\bf Left Panel:} $L_m = p$, {\bf Right Panel:} $L_m = -\rho$.}
    \label{Fig7} 
 \end{figure}
\end{widetext}

Another diagnostic mechanism for the cosmological model is the state finder $(r,s)$ defined as,
$$
r = \frac{\dddot{a}}{aH^3}, 
\quad 
s = \frac{r-1}{3\left(q - \tfrac{1}{2}\right)},
$$

where $r$ and $s$ respectively be the jerk and snap parameter. To note, for $(r,s) = (1,0)$, the model corresponds to $\Lambda$CDM, for $r < 1, s > 0$ it leads to quintessence region and for Chaplygin gas it is, $r > 1, s < 0$. Fig. \ref{Fig8} illustrates the trajectories of the $(r,s)$ pair for both the choice of the Lagrangian, $\mathcal{L}_m =  p$ (left panel) and $\mathcal{L}_m =-\rho$ (right panel), using CC+DES-SN5YR and CC+Pantheon$^+$ datasets. The trajectories of the logarithmic $f(Q,T)$ model, for both Lagrangian prescriptions and datasets, the Universe evolves, the paths deviate into the quintessence region, reflecting an dark energy EoS $\omega > -1$ driving late-time acceleration, and in certain regimes approaching behavior similar to Chaplygin gas (CG) models. This indicates that these models can successfully reproduce late-time cosmic acceleration, while allowing for richer dynamical properties than the standard $\Lambda$CDM cosmology. The differences of evolution between $L_m=p$ and $L_m=-\rho$ are visible in the paths traced in the diagnostic plane, highlighting how the choice of matter Lagrangian influences the detailed dynamical evolution. The deviation from $\Lambda$CDM is slightly more pronounced for $\mathcal{L}_m = -\rho$, indicating a stronger dynamical dark energy behavior compared to $\mathcal{L}_m = p$, which remains closer to the standard trajectory. In both cases, the model supports a quintessence-like scenario.

\begin{widetext}
~\begin{figure}[H]
\centering
\includegraphics[width=18cm]{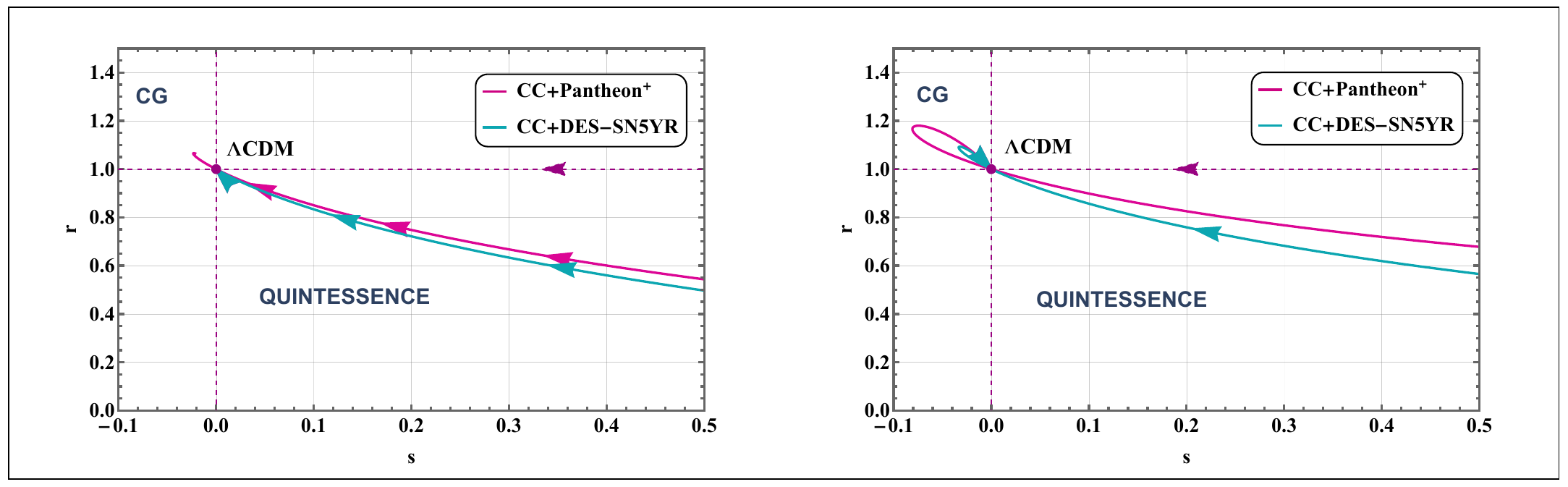}
\caption{Evolution of the Statefinder parameters $\{r,s\}$. {\bf Left Panel:} $L_m = p$, {\bf Right Panel:} $L_m = -\rho$.}
\label{Fig8} 
\end{figure}
\end{widetext}

Now, we shall compute the age of the Universe using the formula,

$$
t_U(z) = \int_z^{z_{\text{ls}}} \frac{dz'}{(1+z')\,H(z')}
$$

where $z_{\text{ls}} = 1089$ corresponds to the redshift of the last scattering surface. The present age of the Universe is then obtained as $t_0 = t_U(0)$, which exhibits an inverse dependence on the Hubble constant $H_0$. Within the logarithmic $f(Q,T)$ model, we obtain $t_0 = 14.383 \, \text{Gyr}$ for the CC+DES-SN5YR dataset and $t_0 = 13.917 \, \text{Gyr}$ for the CC+Pantheon$^+$ dataset when the matter Lagrangian is chosen as $\mathcal{L}_m = p$. For the case $\mathcal{L}_m = -\rho$, the corresponding values are $t_0 = 13.800 \, \text{Gyr}$ (CC+DES-SN5YR) and $t_0 = 13.731 \, \text{Gyr}$ (CC+Pantheon$^+$). These results are consistent with independent observational estimates of the cosmic age. For example, Ref. \cite{Age_Hubble} suggests $t_0 \approx 14.46 \pm 0.8 \, \text{Gyr}$, while the Planck satellite results yield a more precise determination, $t_0 \approx 13.78 \pm 0.02 \, \text{Gyr}$ \cite{Plank_2020}. Stellar evolution studies also support similar values, with $t_0 \approx 13.7 \, \text{Gyr}$ \cite{Creevey_2015}, consistent with more recent refinements \cite{Tang_2021}. Earlier constraints, such as $t_0 \approx 13.8 \pm 4 \, \text{Gyr}$ \cite{Cowan_2002}, also falls within this range.

\begin{widetext}
       ~\begin{figure}[H]
    \centering
    \includegraphics[width=18cm]{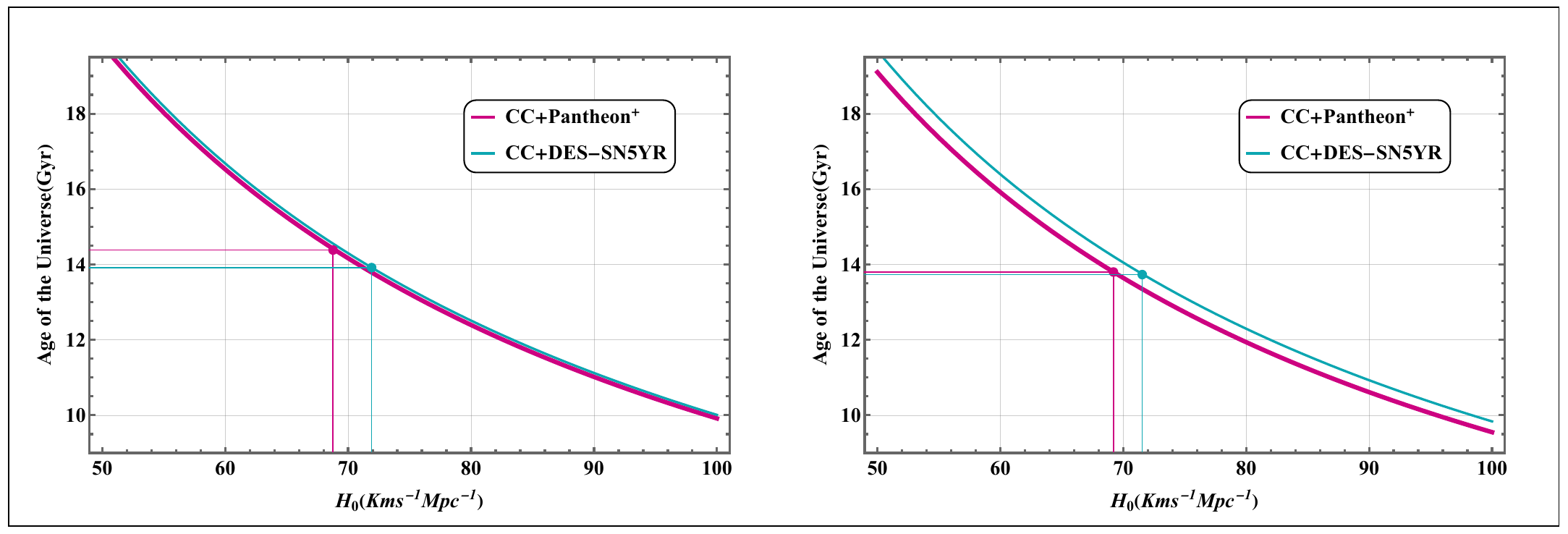}
    \caption{Behavior of the cosmic age $t(z)$. {\bf Left Panel:} $L_m = p$, {\bf Right Panel:} $L_m = -\rho$.}
    \label{Fig9} 
 \end{figure}
 \end{widetext}

\section{ Conclusions}\label{Sec:6} 
We have presented accelerating cosmological models of the Universe in $f(Q,T)$ gravity considering two different matter matter Lagrangian such as $\mathcal{L}_m = p$ and $\mathcal{L}_m = -\rho$. The well motivated logarithmic form of the functional $f(Q,T)$ has been incorporated. Performing the MCMC analysis, the model parameters are constrained using CC dataset and its combination with Pantheon$^+$ and DES-SN5YR datasets. After obtaining the range of the parametric values of the model parameters, we analyzed the dynamical parameters. The effective EoS parameter tends toward $\omega \to -1$, which signals the onset of a dark energy dominated phase.  Additional insights are obtained from the density parameter evolution, which confirms that the total contribution $\Omega_m + \Omega_f \approx 1$, in line with a spatially flat Universe.

Further, we have anlyzed the geometrical parameters to understand the accelerating behavior of the Universe. The evolution of deceleration parameter $q(z)$ confirms the transition from a decelerated to an accelerated regime, with transition redshifts in the range $z_{\text{tr}} \sim 0.64 - 0.77$. At the present epoch, the deceleration parameter approaches negative values ($q_0 = -0.671$ and $q_0 = -0.646$ for $\mathcal{L}_m=p$ and $q_0 = -0.651$ and $q_0 = -0.612$ for $\mathcal{L}_m=-\rho$), reflecting the dominance of cosmic acceleration. The state finder analysis in the $(r,s)$ plane demonstrates that the trajectories remain close to the $\Lambda$CDM fixed point $(r=1,s=0)$ but evolve into the quintessence region at low redshifts, further strengthening the conclusion that the model supports quintessence-like behavior. The behavior of $Om(z)$ further validates this behavior. Moreover, the estimated age of the Universe lies in the range $t_0 \sim 13.7 - 14.3$ Gyr, consistent with results from CMB and stellar evolution studies.

Finally, the results confirm that the logarithmic $f(Q,T)$ model provides a geometrical explanation for the late-time acceleration without requiring an explicit dark energy component. While both choices of the matter Lagrangian yield compatible cosmological behaviors, small quantitative differences arise, with $\mathcal{L}_m = p$ generally predicting slightly larger cosmic ages. This highlights the sensitivity of the model to the underlying matter Lagrangian choice and shows that even within a single modified gravity framework, the specific prescription can influence cosmological predictions.

\section*{Acknowledgement} RB acknowledges the financial support provided by the University Grants Commission (UGC) through Junior Research Fellowship UGC-Ref. No.: 211610028858 to carry out the research work. BM acknowledges the support of IUCAA, Pune (India), through the visiting associateship program.

\bibliographystyle{utphys}
\bibliography{references}

\end{document}